\documentclass{article}

\usepackage[dvips]{graphicx}
\usepackage{bm}

\title{%
  An Equilibrium for Frustrated Quantum Spin Systems\\
in the Stochastic State Selection Method}

\author{%
Tomo  \textsc{Munehisa} and
Yasuko \textsc{Munehisa}
}

\begin{document}
\sloppy
\maketitle
\large

\begin{abstract}
\large

We develop a new method to calculate eigenvalues in frustrated quantum 
spin models.
It is based on the stochastic state selection (SSS) method, 
which is an unconventional Monte Carlo technique we have investigated 
in recent years.
We observe that a kind of equilibrium is realized under some conditions 
when we repeatedly operate a Hamiltonian and a random choice operator, 
which is defined by stochastic variables in the SSS method,
to a trial state.
In this equilibrium, which we call the SSS equilibrium, 
we can evaluate the lowest eigenvalue of the Hamiltonian 
using the statistical average of the normalization factor of the 
generated state.

The SSS equilibrium itself has been already observed in  
unfrustrated models.   
Our study in this paper shows that we can also see the equilibrium 
in frustrated models, with some restriction on values of a parameter 
introduced in the SSS method.
As a concrete example, we employ the spin-1/2 frustrated $J_1$-$J_2$ 
Heisenberg model on the square lattice. 
We present numerical results on the 20-, 32-, 36-site systems, 
which demonstrate that statistical averages 
of the normalization factors reproduce the known exact eigenvalue 
in good precision. 
Finally we apply the method to the 40-site system.
Then we obtain the value of the lowest energy eigenvalue with an error 
less than $0.2 \%$. 

\end{abstract}

\section{Introduction}
\label{intro}

The quantum Monte Carlo method is well established in the 
numerical study of quantum spin systems. 
This method has given us fruitful results to understand properties of 
unfrustrated systems, especially of the spin-1/2 quantum
Heisenberg anti-ferromagnet on bipartite lattices \cite{rsh}.
But, due to the so-called sign problem, 
the method is not useful for studies of frustrated systems.
In contrast to this the exact diagonalization is applicable even when 
systems are frustrated. 
By this method, however, one can only deal with systems on small-sized 
lattices.
In order to overcome this difficulty lots of active studies for 
numerical methods have been made in the Monte Carlo approach.
Among them the re-normalization group method \cite{ki} and the reconfiguration
method combined with the fixed node method \cite{sor} are quite noticeable. 
Also, the extensions of the density matrix re-normalization group 
method \cite{white} is worth noting. 

Recently we have developed a new Monte Carlo method to evaluate energy 
eigenvalues of quantum spin systems \cite{mune,mune2,mune3,mune4,mune5}.
We call it the {\em stochastic state selection} (SSS) method \cite{footsss}.
This method provides an essential improvement on early attempts 
to truncate the Hilbert space by omitting small 
values of coefficients \cite{omits}.
{\em Not} being based on importance samplings, the method 
enables us to select a relatively small number of states from a
vast full vector space.
This selection is simple and is mathematically justified
so that one can calculate correct values of inner products
through statistical averaging processes. 
In order to realize this selection,
we employ an operator which we call the {\em random choice operator}. 
This operator is represented by a diagonal matrix whose elements
are stochastic variables with unit averages.

It is possible to combine the SSS method with various techniques for 
numerical studies.
For instance, several applications with the power method are in 
refs. \cite{mune2, mune3}.
By repeating alternate operations of the random choice operator
and the Hamiltonian to a trial state, 
we numerically demonstrated that we can obtain expectation values for 
powers of the Hamiltonian even when limited computer memory resources 
are available. 
We also have combined the SSS method with the Lanczos method in order to 
obtain the lowest energy eigenvalue of the spin-1/2 quantum Heisenberg 
anti-ferromagnet on the triangular lattice up to 48 sites \cite{mune5}.
It should be noted that these applications of the SSS method are simple 
from mathematical points of view, because the statistical averaging 
process for random variables is clear.

In this paper we discuss another application of the SSS method that is 
essentially different from the applications stated above.
Here we consider a number of intermediate states which are successively 
generated by the random choice operator and an operator related to 
the Hamiltonian, and measure their normalization factors.
Generally the standard deviation of the inner product between the 
initial trial state and 
the $L$-th generated state increases as $L$ grows.
In some conditions, however, the deviation does not
increase but become constant when $L$ goes beyond some value.
We call this phenomenon as an {\em equilibrium}, because in this
situation statistical averages of the inner product  
do not depend on $L$ after suitable normalizations.
This type of equilibrium is quite interesting since from values of 
the normalization 
factor one can obtain the lowest energy eigenvalue precisely and simply.
We first observed such phenomena in our study of the $J_1$-$J_2$ Heisenberg 
model with unfrustrated \cite{footuf} couplings \cite{mune4}.
In this paper we study the equilibrium in the frustrated case.
From qualitative discussions we show that the equilibrium for 
frustrated systems exists in some parameter region. 
A concrete example for numerical examinations is 
the spin-1/2 $J_1$-$J_2$ Heisenberg model on the square lattice
with couplings $J_1=1$ and $J_2 = 0.5$, the model which is one of the 
most popular frustrated models with ample numerical 
results \cite{szp,cs,cbps,cfrt,kosw}.
Studying this model up to 40 sites, we find that our results give 
tangible evidences for the equilibrium.   

This paper is organized as follows. Next two sections are devoted 
to definitions and qualitative discussions. 
In Section \ref{ssseql}, after brief descriptions for the SSS method, a definition 
of the SSS equilibrium is given. Here we also explain why the 
equilibrium enables us to calculate the lowest energy eigenvalue. 
Section \ref{exssseql} is for summarized discussions on where the equilibrium 
takes place. 
We briefly repeat our previous argument \cite{mune4} with which we 
concluded that the equilibrium should appear in any unfrustrated 
system. Then we argue that 
for frustrated systems the equilibrium also exists in some parameter region.
In Section \ref{nstudy} we present numerical results on the frustrated 
$J_1$-$J_2$ Heisenberg model.
We first direct our attention to the model on a 20-site lattice. 
On this small lattice it is easy to obtain the exact ground state by 
the diagonalization method. 
Consequently we can directly measure overlaps between the exact ground 
state and the states generated through stochastic state selections. 
Note that, as is discussed in Section \ref{ssseql}, those overlaps are most 
effective to prove the existence of the SSS equilibrium.  
In the 20-site case it is also possible to make detailed observations 
on several quantities introduced in Section \ref{exssseql}.   
In Subsection \ref{ns20} we will show that assumptions we employ in 
Section \ref{exssseql} are mostly reasonable for the model and that 
the eigenvalue we obtain through normalization factors is in good agreement
with the exact ground-state energy. 
We also examine how many basis states are needed in our method to 
calculate the eigenvalue with enough precision.
Subsection \ref{ns32and36} is to present our results for the 32- and 36-site systems. 
For these sizes we construct bases taking some symmetries into account. 
Without knowing the exact ground state, we use not overlaps but normalization 
factors to observe the SSS equilibrium in these systems. 
Our results indicate that the eigenvalues we calculate in the
equilibrium nicely 
reproduce the exact ground-state energies reported in ref. \cite{szp}. 
In Subsection \ref{ns40} we present results for the 40-site system.
The final section is devoted to summary and discussions.

\section{Stochastic State Selection Equilibrium}
\label{ssseql}

First we briefly review the SSS method \cite{mune3}.
Consider a system whose Hamiltonian is $\hat H$ and suppose its full 
vector space is given by a basis 
$\{ \mid i \rangle \}$ ($i=1,...,N_{\rm V}$).
Here $N_{\rm V}$ denotes the size of the full vector space.
We denote the lowest energy eigenvalue of $\hat H$ by $E$ and its 
eigenstate by $\mid \psi_{\rm E} \rangle $.
For convenience a new notation $\hat Q$ is used hereafter,
\begin{eqnarray}
\hat Q \equiv l\hat I - H \ ,
\label{qdef}
\end{eqnarray}
with the identity operator $\hat I$ and a positive number $l$.
Since the present method is based on the power method,  
the value of $l$ should be chosen to ensure that $Q \equiv l-E (>0)$ 
is the largest among absolute values of eigenvalues for $\hat Q$.  
    
Let us employ a normalized trial state
$\mid \psi^{(0)}\rangle \equiv \sum_{i} \mid i \rangle c_i^{(0)}$,  
$\langle \psi^{(0)} \mid \psi^{(0)}\rangle =1$.  
We introduce random choice operators
\begin{eqnarray}
\hat{M}_{\{ \eta ^{(m)}\}} =\sum_{i=1}^{N_{\rm V}}
  \mid i \rangle  \eta_i^{(m)} \langle i \mid \ \ \ (m=1,2,\cdots) \ , 
\label{mdef}
\end{eqnarray} 
in order to calculate 
\begin{eqnarray}
\hat Q \hat{M}_{\{ \eta ^{(L)}\}} \hat Q \hat{M}_{\{ \eta ^{(L-1)}\}} 
\cdots \hat Q \hat{M}_{\{ \eta ^{(1)}\}} \mid  \psi^{(0)} \rangle  
\label{ldef}
\end{eqnarray}
instead of $ \hat Q ^L \mid  \psi^{(0)} \rangle $, 
the state which would become close to 
$(l-E)^L \mid  \psi_{\rm E} \rangle$ for large values of $L$.  
The random variable $\eta_i^{(m)}$ in (\ref{mdef}) is generated following 
the {\em on-off probability function} $ P_i^{(m)}(\eta)$.  
Using a positive parameter $\epsilon$ and 
the coefficient $c_i^{(m-1)}$ in the normalized intermediate state  
$\mid \psi^{(m-1)} \rangle \equiv \sum \mid i \rangle c_i^{(m-1)}$,
which is proportional to $\hat Q \hat{M}_{\{ \eta^{(m-1)}\}} \cdots \hat Q
\hat{M}_{\{ \eta^{(1)}\}} \mid  \psi^{(0)} \rangle$, we define
\begin{eqnarray}
 P_i^{(m)}(\eta) \equiv \frac{1}{a_i ^{(m)}} 
\delta \left( \eta - a_i ^{(m)} \right)
+ \left( 1 - \frac{1}{a_i ^{(m)}} \right) \delta \left( \eta \right) \ , 
\ \ \ \ \ \ \frac{1}{a_i ^{(m)}} 
\equiv \min \left(1,\frac{|c_i^{(m-1)}|}{\epsilon} \right) \ .
\label{pidefx}
\end{eqnarray}
Then starting from a given $\mid \psi^{(0)} \rangle$ we sequentially calculate 
\begin{eqnarray}
\mid \psi^{(m)} \rangle \equiv
\hat Q \hat{M}_{\{\eta ^{(m)} \}} \mid \psi^{(m-1)}\rangle / C^{(m)} 
\label{psim} 
\end{eqnarray}
for $m=1,2,3,\cdots$, where $ C^{(m)} \ (>0)$ is the normalization 
factor determined by
\begin{eqnarray}
\left[ C^{(m)} \right] ^2 &=& 
\langle \psi^{(m-1)} \mid  \hat{M}_{\{\eta ^{(m)} \}}
 \hat Q ^2 \hat{M}_{\{\eta ^{(m)} \}} \mid \psi^{(m-1)} \rangle \ .
\label{cmsquare}
\end{eqnarray}
We also define a {\em random state} $\mid \chi ^{(m)} \rangle g^{(m)}$ 
with the normalization condition 
$\langle \chi^{(m)} \mid \chi^{(m)} \rangle = 1$ by 
\begin{eqnarray}
\mid \chi^{(m)} \rangle \ g^{(m)} \equiv \hat{M}_{\{ \eta ^{(m+1)} \}}
\mid \psi^{(m)} \rangle \ - \mid \psi^{(m)} \rangle = 
\sum_i \mid i \rangle c_i^{(m)} \left(\eta_i^{(m+1)} - 1 \right) \ .
\label{chim}
\end{eqnarray}
It should be kept in mind that for any state 
$\mid \Phi \rangle \equiv \sum \mid i \rangle b_i$ 
with $b_i$'s irrelevant to $\{\eta_i^{(m+1)}\}$ 
the statistical average of the inner product between $\mid \Phi \rangle$
and this random state is zero,
\begin{eqnarray}
\langle \! \langle \ \langle \Phi \mid
\chi^{(m)} \rangle g^{(m)} \ \rangle \! \rangle = 0 \ , 
\label{sinn0}
\end{eqnarray}
because  
\begin{eqnarray}
\langle \! \langle \eta_i^{(m+1)} \rangle \! \rangle  
\equiv \int_0^{\infty} \eta_i^{(m+1)} P_i^{(m+1)}(\eta_i^{(m+1)}) 
d \eta_i^{(m+1)} = 1 \ , 
\label{etaimsa}
\end{eqnarray}
from (\ref{pidefx}) \cite{footeta}.
Also remember that 
\begin{eqnarray}
\langle \! \langle \left[ g^{(m)}\right]^2 \rangle \! \rangle &=&
\langle \! \langle \sum_i \left[ c_i^{(m)}\right]^2
\left(\eta_i^{(m+1)}-1 \right)^2 \rangle \! \rangle = 
\sum_{0 < |c_i^{(m)}| < \epsilon } \left[ c_i^{(m)}\right]^2 
\left(\frac{\epsilon}{| c_i^{(m)}|} -1 \right)\nonumber \\
&=& \epsilon \sum_{|c_i^{(m)}| < \epsilon }|c_i^{(m)}| \ 
- \sum_{ |c_i^{(m)}| < \epsilon }\left[ c_i^{(m)}\right]^2 \ , 
\label{gmsquare}  
\end{eqnarray}
where we use (\ref{etaimsa}) and 
\begin{eqnarray}
\langle \! \langle \left[ \eta_i^{(m+1)} \right]^2 \rangle \! \rangle  
\equiv 
\int_0^{\infty} \left[\eta_i^{(m+1)}\right]^2
 P_i^{(m+1)}(\eta_i^{(m+1)}) d \eta_i^{(m+1)} = a_i^{(m+1)} \ .
\label{etaim2sa}
\end{eqnarray}

Now we define the SSS equilibrium \cite{footequil}. 
We divide the intermediate state $\mid \psi^{(m)} \rangle $ into
a part which is proportional to $\mid \psi_{\rm E} \rangle $ and the
rest, 
\begin{eqnarray}
\mid \psi^{(m)} \rangle = \mid \psi_{\rm E} \rangle w^{(m)} + 
\mid \zeta ^{(m)} \rangle s^{(m)} \ . 
\label{sepes}
\end{eqnarray}
Here 
\begin{eqnarray}
 w^{(m)} &\equiv& \langle \psi_{\rm E} \mid \psi^{(m)} \rangle 
\label{defwm} 
\end{eqnarray}
and
\begin{eqnarray}
\mid \zeta ^{(m)} \rangle s^{(m)} &\equiv&  \mid \psi^{(m)} \rangle -
  \mid \psi_{\rm E} \rangle w^{(m)}  
\label{defzmsm}
\end{eqnarray}
with the normalization condition 
$\langle \zeta ^{(m)} \mid \zeta ^{(m)} \rangle =1$.
Note that 
\begin{eqnarray}
\langle \psi_{\rm E}\mid \zeta ^{(m)} \rangle = 0  
\label{pezm}
\end{eqnarray}
and 
\begin{eqnarray}
\left[w^{(m)}\right]^2 + \left[s^{(m)}\right]^2 = 1   
\label{wm2sm2}
\end{eqnarray}
by definition.
What we mean by the SSS equilibrium is that 
there exists a limit $w^{({\rm eq})}$ defined by      
\begin{eqnarray}
\lim_{m_{\rm t} \rightarrow \infty} \frac{1}{m_{\rm t}} 
\sum_{m=m_{\rm s}}^{m_{\rm s}+m_{\rm t}-1}w^{(m)}= w^{({\rm eq})} \ , 
\ \ \ \ \ (0 < w^{({\rm eq})} \leq 1) \ , 
\label{wlim}
\end{eqnarray}
where $w^{({\rm eq})}$ is independent of $m_{\rm s}$ whenever 
$m_{\rm s}$ is greater than or equal to some value of $m$.

Finally let us comment on a relation useful in the equilibrium 
in order to extract the value of $E= l - Q$.
Using (\ref{psim}), (\ref{chim}) and (\ref{defwm}) we obtain 
\begin{eqnarray}
w^{(m+1)}  C^{(m+1)} &=& \langle \psi_{\rm E} \mid \psi^{(m+1)} 
\rangle \ C^{(m+1)} \nonumber\\
&=& \langle \psi_{\rm E} \mid  \hat Q \hat M_{\{\eta ^{(m+1)} \}} 
\mid \psi^{(m)} \rangle \nonumber \\
&=& \langle \psi_{\rm E} \mid
\hat Q \ \{ \ \mid \psi^{(m)} \rangle + \mid \chi^{(m)} \rangle g^{(m)} \  \}
\nonumber \\
&=& Q \{ \  \langle \psi_{\rm E} \mid \psi^{(m)} \rangle
+ \langle \psi_{\rm E} \mid \chi^{(m)} \rangle g^{(m)} \  \}
\nonumber \\
&=& Q  w^{(m)} + Q \langle \psi_{\rm E} \mid \chi^{(m)} \rangle g^{(m)} \ .  
\label{psiwmp1}
\end{eqnarray}
If the second term in the right-hand side is negligible, it leads to 
\begin{eqnarray}
w^{(m+1)} \simeq \frac{Q}{ C^{(m+1)}} \  w^{(m)} \ .
\label{wmp1sim}
\end{eqnarray}
Then, in the equilibrium where 
$w^{(m)} \simeq w^{(m+1)} \simeq w^{({\rm eq})} \neq 0$ 
for sufficiently large values of $m$, we can expect 
\begin{eqnarray}
Q \simeq C^{(m+1)} \ .  
\label{estime} 
\end{eqnarray}
Thus we become aware that the value of $E$ can be estimated 
from the normalization factor $C^{(m+1)}$.

\section{Existence of the SSS Equilibrium}
\label{exssseql}

In this section we  present analytic and qualitative discussions on
the existence of the SSS equilibrium.
This section consists of three subsections. 
The first subsection gives some equations for $C^{(m+1)} $
assuming that fluctuations can be neglected.
Next subsection is devoted to summarized discussions for 
unfrustrated systems, where all elements of the operator $\hat Q$ are 
non-negative \cite{mune4}. 
In the third subsection we show that the SSS
equilibrium also exists for frustrated systems when the parameter
$\epsilon $ is small enough. 

\subsection{An Equation for the Equilibrium}
\label{suba}

We pay our attention to a relation led from (\ref{psim}), (\ref{chim}) 
and (\ref{sepes}),
\begin{eqnarray}
\mid \psi^{(m+1)} \rangle  C^{(m+1)} &=& 
\hat Q \hat M _{\{\eta^{(m+1)}\} } \mid \psi ^{(m)} \rangle \nonumber \\ 
&=& Q \mid \psi_{\rm E} \rangle w^{(m)} + 
\hat Q \mid \zeta^{(m)} \rangle s^{(m)}+\hat Q \mid \chi^{(m)} \rangle 
g^{(m)} \ .
\label{psicmp1}
\end{eqnarray}
With normalization conditions and (\ref{pezm}) and (\ref{wm2sm2}) it yields
\begin{eqnarray}
\left[ C^{(m+1)}\right]^2 &=& Q^2  \left[ w^{(m)}\right]^2 \nonumber \\
&+& \langle \zeta^{(m)}\mid \hat Q ^2 \mid \zeta^{(m)} \rangle
\left(1- \left[ w^{(m)}\right]^2 \right)
+ \langle \chi^{(m)}\mid \hat Q^2 \mid \chi^{(m)} \rangle
\left[ g^{(m)}\right]^2 \nonumber \\
&+& 2 Q^2 w^{(m)} \langle \psi_{\rm E} \mid \chi^{(m)} \rangle g^{(m)}
+ 2 s^{(m)} \langle \zeta^{(m)}  \mid \hat Q^2 \mid \chi^{(m)} \rangle 
g^{(m)} \ .
\label{cmp1sq}
\end{eqnarray}
We assume that both 
$\langle \zeta^{(m)} \mid \hat Q^2 \mid \zeta^{(m)} \rangle$ and
$\langle \chi^{(m)} \mid \hat Q^2 \mid \chi^{(m)} \rangle$ are
independent of $m$ when $m$ is large enough,
 \begin{eqnarray}
\langle \zeta^{(m)} \mid \hat Q^2 \mid \zeta^{(m)} \rangle \ &\simeq& \
Q_{2\zeta} \ , \label{as1} \\
\langle \chi^{(m)} \mid \hat Q^2 \mid \chi^{(m)} \rangle \ &\simeq& \
Q_{2\chi} \ , \label{as2}
\end{eqnarray}
where $Q_{2\zeta}$ and $Q_{2\chi}$ denote positive constants. 
It should be noted that both $Q_{2\zeta}$ and $Q_{2\chi}$ are always 
less than $Q^2$ because 
$Q^2 = \langle \psi_{\rm E} \mid \hat Q ^2 \mid \psi_{\rm E} \rangle $ 
is the largest one among $\langle \psi \mid \hat Q ^2 \mid \psi \rangle $ 
by definition.
We also assume both of cross terms in (\ref{cmp1sq}) are negligible, 
notifying that statistical averages of them vanish. 
Thus we obtain a relation
\begin{eqnarray}
\left[ C^{(m+1)}\right]^2 \ \simeq \
Q^2 \left[w^{(m)}\right]^2 + Q_{2\zeta}\left( 1- \left[w^{(m)}\right]^2\right)
+Q_{2\chi} \left[ g^{(m)}\right]^2 \ .
\label{wright}
\end{eqnarray}
Let us here emphasize that $\left[g^{(m)}\right]^2$ measures the 
degree of the difference between $\mid \psi ^{(m)} \rangle$ and 
$\hat M_{\{ \eta^{(m+1)}\ \}} \mid \psi ^{(m)} \rangle$.
We would like to notify again that the statistical average of 
$\left[g^{(m)}\right]^2$ is calculated from coefficients for 
$\mid \psi ^{(m)} \rangle $ and the parameter $\epsilon$ 
(see (\ref{gmsquare})). 

\subsection{The Equilibrium in Unfrustrated Systems}
\label{subb}

Here we briefly comment how we conclude that the SSS equilibrium 
should exist in unfrustrated cases with any value of $\epsilon$. 
Detailed discussions and several numerical examinations with the 
$J_1$-$J_2$ Heisenberg model ($J_2/J_1 =0, \ -1$) are presented in 
ref. \cite{mune4}.
In this subsection we limit ourselves to the case 
$\mid \psi^{(0)} \rangle = \mid \psi_{\rm E} \rangle$ 
in order to make our analysis clear.
Discussions for a good approximate initial trial state can be made 
in a similar manner.  
  
For unfrustrated systems it is always possible to choose an adequate basis
$\{\mid i \rangle \}$ for which all $f_i$'s in 
$\mid \psi_{\rm E} \rangle \equiv \sum_i \mid i \rangle f_i$ as well as
$q_{ij} \equiv \langle i \mid \hat Q \mid j \rangle$'s are non-negative.
Note that all coefficients in the expansion of
$\mid \psi ^{(1)} \rangle$,  $\mid \psi ^{(2)} \rangle$, $\cdots$ are
also non-negative then. 
This is because, in the relation we learn from (\ref{psim}),
\begin{eqnarray}
c_i^{(m)} = \sum_j  q_{ij} c_j^{(m-1)} \eta_j^{(m)} / C^{(m)} \ ,
\label{cim}
\end{eqnarray}
$q_{ij} \geq 0$ for all $i$ and $j$
and $\eta_j^{(m)} /C^{(m)} \geq 0$ for all $j$ by definition.
Let us then examine the first term in (\ref{gmsquare}). 
One upper bound for it is given by 
\begin{eqnarray}
\epsilon \sum_{|c_i^{(m)}| < \epsilon }|c_i^{(m)}| \ \leq \
\epsilon \sum_i |c_i^{(m)}| = \epsilon \sum_i c_i^{(m)} \ ,
\label{gm1sup}
\end{eqnarray}
where the last equality follows from the fact that all 
$c_i^{(m)}$ are non-negative here.
Further, we can expect that, in positive definite cases,  
\begin{eqnarray}
\epsilon \sum_i c_i^{(m)} \sim w^{(m)} \epsilon \sum_i f_i
\label{ftermsim}
\end{eqnarray}
holds for any $m$ \cite{mune4}.   
Now we add one more assumption in order to clearly propose 
a relation to be expected in the SSS equilibrium.
Noting that $\sum_{m} \left[c_i^{(m)} \right]^2 =1$, 
we assume for the second term of (\ref{gmsquare}) that
\begin{eqnarray}
\sum_{|c_i^{(m)}| < \epsilon } \left[c_i^{(m)} \right]^2 \ &\simeq& \ K \  
\label{as3} 
\end{eqnarray}
holds with sufficiently large values of $m$,
where $K$ $(0 < K \leq 1)$ is a constant defined by $\epsilon$.
This brings, together with (\ref{ftermsim}), a relation  
\begin{eqnarray}
\left[ g^{(m)} \right]^2 \ &\simeq& \ G w^{(m)} - K \ , \ \ \ \ \ G \equiv
\epsilon \sum_i f_i  \ .
\label{as4}
\end{eqnarray} 
Combining (\ref{wmp1sim}), (\ref{wright}) and (\ref{as4}), we obtain a 
recursive relation for $w^{(m)}$, 
\begin{eqnarray}
w^{(m+1)} \simeq \frac{Q w^{(m)}} {\sqrt{(Q^2 -Q_{2\zeta}) 
\left[ w^{(m)} \right] ^2 
+ G Q_{2\chi} w^{(m)} + (Q_{2\zeta} - K Q_{2\chi}) }} \ .
\label{wmrecu}
\end{eqnarray}
A simple mathematical analysis on (\ref{wmrecu}) \cite{mune4} 
leads us to the conclusion that 
$w^{({\rm eq})}$ exists as far as $G > K$ and 
$Q_{2\zeta} > K Q_{2\chi}$ \cite{footcond}.  
Also it is concluded that $w^{({\rm eq})}$ should satisfy 
a quadratic equation for $w$,
\begin{eqnarray}
\left(Q^2- Q_{2\zeta}\right) w^2 + G Q_{2\chi}  w -
\left(Q^2 - Q_{2\zeta} + K Q_{2\chi} \right) = 0 \ .
\label{equcw}
\end{eqnarray}
The relevant solution which belongs to the interval $(0,1]$ is 
\begin{eqnarray}
w^{({\rm eq})} = -q + \sqrt{q^2+1+\kappa} \ , \ \ \ q \equiv 
\frac{1}{2} \cdot
\frac{G Q_{2\chi}}{Q^2-Q_{2\zeta}} \ (> 0) \ , \ \ \ \kappa \equiv 
\frac{K Q_{2\chi}}{Q^2-Q_{2\zeta}} \ (> 0) \ .
\label{wsol}
\end{eqnarray} 
Thus we come to a conclusion that the SSS equilibrium exists 
in unfrustrated systems. 

\subsection{The Equilibrium in Frustrated Systems}
\label{subc}

In this subsection we show that the SSS equilibrium should, at least 
for small values of $\epsilon$, also exist for frustrated systems.

Let us find an equation for $w^{({\rm eq})}$ in frustrated systems 
from (\ref{wright}).
We again start from examining $\left[ g^{(m)} \right] ^2$, 
with an assumption that it becomes almost constant ($\equiv g^2$) 
for sufficiently large values of $m$. 
Then (\ref{gmsquare}) brings an upper bound,
\begin{eqnarray}
g^2 \simeq \left[ g^{(m)}\right]^2  \simeq 
\langle \! \langle \left[ g^{(m)}\right]^2 \rangle \! \rangle 
\ \leq \ \epsilon \sum_{|c_i^{(m)}| \  < \ \epsilon }|c_i^{(m)}| \ \leq \
\epsilon \sum_{i=1}^{N_{\rm V}}|c_i^{(m)}| \ \leq \ 
\epsilon \sqrt{N_{\rm V}} \ .
\label{g2uplim}
\end{eqnarray}   
The last inequality is based on the fact that 
$\displaystyle \max_{\sum c_i^2 =1} {\sum |c_i|}= \sqrt{N_{\rm V}}$.
 
If the SSS equilibrium exists, $w^{({\rm eq})}$ should satisfy 
the following equation for $w$, 
\begin{eqnarray}
\left( Q^2 - Q_{2\zeta} \right) \left( 1- w^2 \right) \ 
\simeq \ Q_{2\chi} g^2 \ ,
\label{weq0}
\end{eqnarray}
which we obtain using (\ref{estime}) and (\ref{wright}). 
Clearly this equation, where $Q^2 - Q_{2\zeta} > 0$ and $Q_{2\chi} \geq 0$ 
always hold by definition,
has a relevant solution if $ Q_{2\chi} g^2$ is less than 
$ Q^2 - Q_{2\zeta}$.  
Because of the upper bound (\ref{g2uplim}) we then see that there 
exists a solution   
\begin{eqnarray}
w^{({\rm eq})} = \sqrt{1-\frac{Q_{2\chi} g^2}{Q^2-Q_{2\zeta}}} \ , 
\ \ \ (0 < w^{({\rm eq})} \leq 1 ) 
\label{weq0sol}
\end{eqnarray}
for small values of $\epsilon$. 
Therefore we conclude that the SSS equilibrium is realized 
in frustrated systems when the value of the parameter $\epsilon$ 
is small enough. 

\section{Numerical Study}
\label{nstudy}
 
Now we present numerical results on the frustrated $J_1$-$J_2$ Heisenberg model. 
The Hamiltonian of the model is 
\begin{eqnarray}
\hat H _{J_1J_2} \equiv J_1 \sum_{(nn)} \bm{S}_i \cdot \bm{S}_j +
J_2 \sum_{(nnn)} \bm{S}_i \cdot \bm{S}_{j'} \ .
\label{hj1j2} 
\end{eqnarray}
Here $\bm{S}_k$ denotes the spin 1/2 operator on the site $k$ and 
summations run over the nearest neighbor pairs $(nn)$ or 
over the next-nearest neighbor pairs $(nnn)$ of the square lattice. 
In this work we restrict ourselves to the $S_z = 0$ sector, where $S_z$ denotes 
the $z$ component of the total spin \cite{footsz}. 
Values of couplings are fixed to be $J_1=1$ and $J_2=0.5$ 
throughout this paper.  

As is mentioned in Section \ref{ssseql}, 
the value of $l$ in the operator $\hat Q \equiv l \hat I - \hat H _{J_1J_2}$   
should be determined to make the largest 
eigenvalue for $\hat Q$ correspond to the lowest eigenvalue of 
$\hat H_{J_1J_2}$, which we denote by $E$. 
Since the largest eigenvalue of $\hat H _{J_1J_2}$ on an 
$N_s$-site lattice is $2(J_1+J_2) N_s /4$ with the largest $S_z (= N_s /2)$, 
this means a condition $2l > (J_1+J_2) N_s /2 - |E|$.
Also keep in mind that very large value of $l$ should be avoided 
because it causes slow convergence. 
Values of $l$ thus chosen for each lattice size with a guess for $E$
will be given in each subsection.  

This section, which includes three subsections,
is for numerical studies on lattices up to $N_s=40$. 
In the first subsection we make a detailed study of the SSS equilibrium 
on a $N_s=20$ lattice. Comparing our results with those obtained 
from the exact eigenstate, we demonstrate that our assumptions 
described in the previous section are reasonable. 
In the second subsection we evaluate the lowest energy eigenvalue on 
$N_s=32$ and 36 lattices, imposing some symmetries on their bases. 
We will see that 
our results are in good agreement with the known exact ground-state energy. 
The third subsection is for the $N_s=40$ results.

\subsection{$N_s$=20 Results}
\label{ns20}

For this lattice size there are $_{20}$C$_{10}$=184,756 configurations 
which fulfill the condition $S_z = 0$. 
We use these configurations as basis states $\mid i \rangle$, 
without assuming any symmetry on the state $\mid \psi^{(m)} \rangle$. 
Let $\hat Q \equiv 3 \hat I - \hat H _{J_1J_2}$ in this subsection.
It is easy to carry out the exact diagonalization of the matrix 
$[\langle i \mid \hat Q \mid j \rangle]$ and find the exact eigenstate  
$\mid \psi _{\rm E} \rangle $. 
Remember that 
$\hat Q \mid \psi _{\rm E} \rangle = Q \mid \psi _{\rm E} \rangle$ and
$Q$ is the eigenvalue for which the value of $|Q|$ is the largest 
among all eigenvalues for $\hat Q$. 
Numerically we obtain $Q=13.0123$, which certainly reproduces 
the exact ground-state energy of the system.    

\begin{figure}[p]
\begin{center}
\scalebox{0.5}{\includegraphics{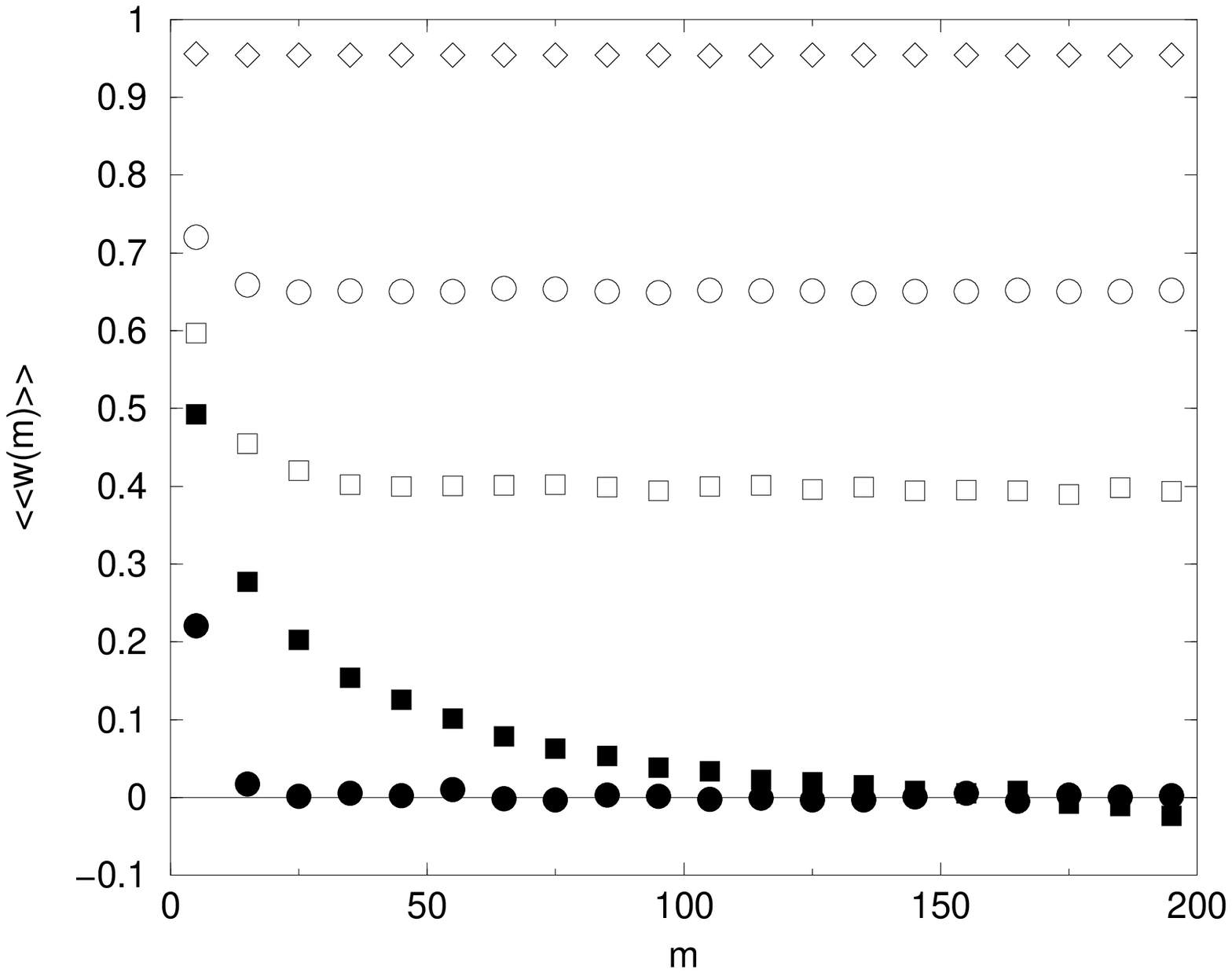}}
\caption{Statistical averages for $w^{(m)}$ on the 20-site lattice 
with $\epsilon =$ 0.05 (filled circles), 0.025 (filled squares), 
0.02 (open squares), 0.015 (open circles) and 0.005 (open diamonds).}
\label{fig:wm20}
\end{center}
\end{figure}

\begin{figure}[h]
\begin{center}
\scalebox{0.5}{\includegraphics{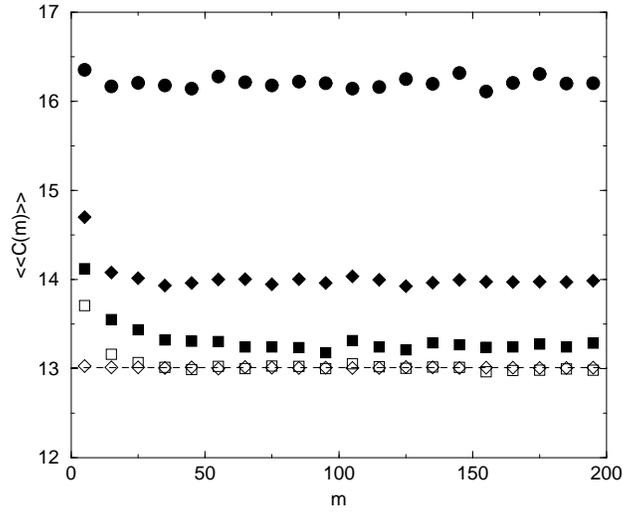}}
\caption{Statistical averages for $C^{(m)}$ on the 20-site lattice 
with $\epsilon = $ 0.05 (filled circles), 0.03 (filled diamonds), 
0.025 (filled squares), 0.02 (open squares) and 0.005 (open diamonds). 
The dashed line indicates the eigenvalue $Q(=3-E)=13.0123$.}
\label{fig:cm20}
\end{center}
\end{figure}

\begin{figure}[p]
\begin{center}
\scalebox{0.5}{\includegraphics{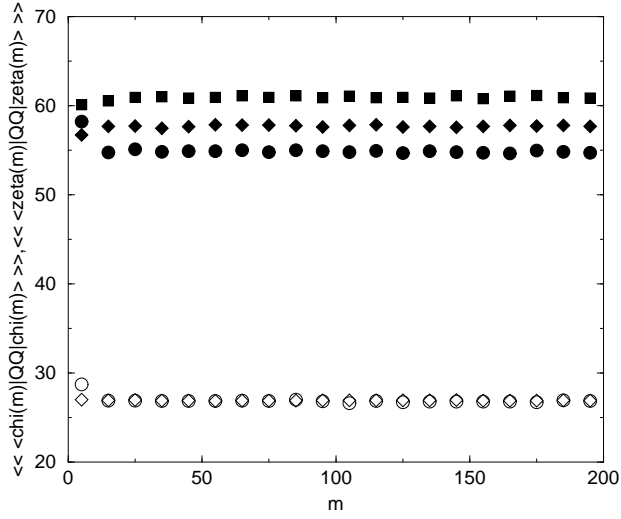}}
\caption{Statistical averages for $\langle \zeta^{(m)} \mid \hat Q ^2 \mid \zeta^{(m)} \rangle$ (filled marks) and 
$\langle \chi^{(m)} \mid \hat Q ^2 \mid \chi^{(m)} \rangle$ (open marks)
on the 20-site lattice.
Circles, squares and diamonds show the data with 
$\epsilon = 0.05$, 0.02 and 0.005, respectively.}
\label{fig:qq20}
\end{center}
\end{figure}

\begin{figure}[h]
\begin{center}
\scalebox{0.5}{\includegraphics{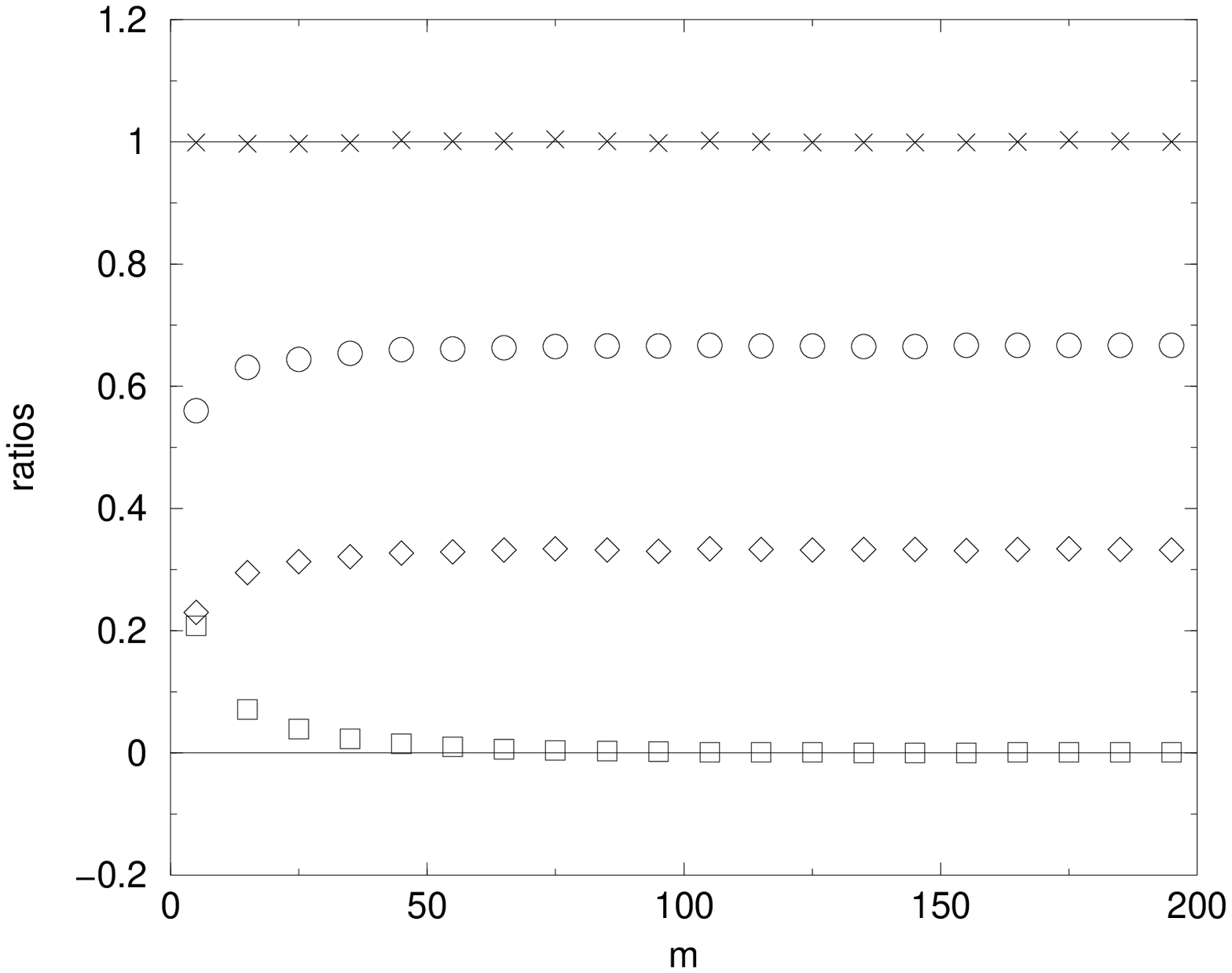}}
\caption{Ratios  
$Q^2\left[ w^{(m)} \right] ^2 / \left[ C^{(m+1)} \right] ^2 $ (squares), 
$Q_{2\zeta} \left( 1 - \left[ w^{(m)} \right] ^2 \right) / 
\left[ C^{(m+1)} \right] ^2 $ (diamonds), 
$Q_{2\chi} \left[ g^{(m)} \right] ^2 / \left[ C^{(m+1)} \right] ^2 $ (circles) and
$\left\{ Q^2\left[ w^{(m)} \right] ^2 +  
Q_{2\zeta} \left( 1 - \left[ w^{(m)} \right] ^2 \right) +
Q_{2\chi} \left[ g^{(m)} \right] ^2
\right\} / \left[ C^{(m+1)} \right] ^2 $ (crosses) 
on the 20-site lattice with $\epsilon = 0.025$.}
\label{fig:rat20a}
\end{center}
\end{figure}

\begin{figure}[p]
\begin{center}
\scalebox{0.5}{\includegraphics{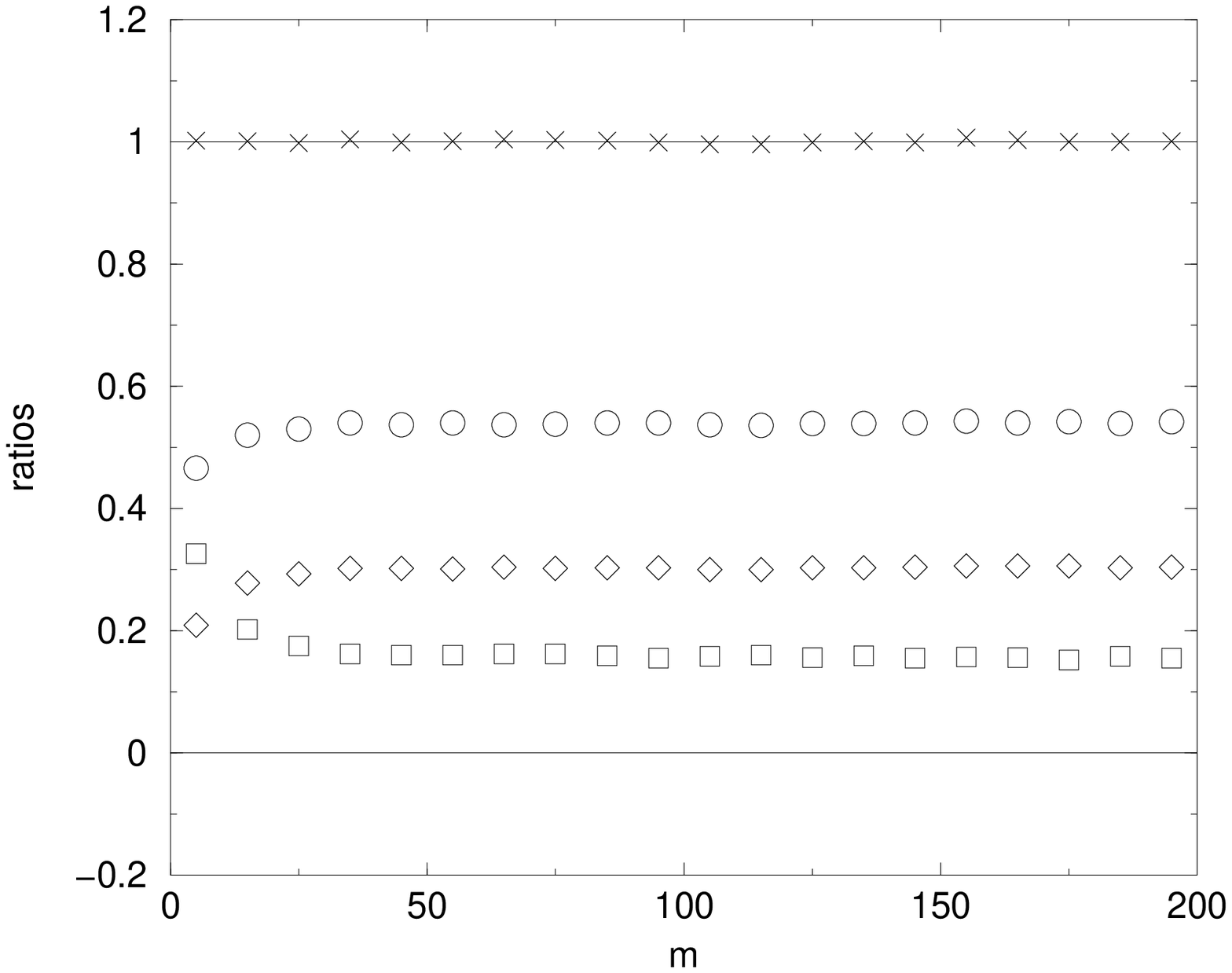}}
\caption{Ratios  
$Q^2\left[ w^{(m)} \right] ^2 / \left[ C^{(m+1)} \right] ^2 $ (squares), 
$Q_{2\zeta} \left( 1 - \left[ w^{(m)} \right] ^2 \right) / 
\left[ C^{(m+1)} \right] ^2 $ (diamonds), 
$Q_{2\chi} \left[ g^{(m)} \right] ^2 / \left[ C^{(m+1)} \right] ^2 $ (circles) and
$\left\{ Q^2\left[ w^{(m)} \right] ^2 +  
Q_{2\zeta} \left( 1 - \left[ w^{(m)} \right] ^2 \right) +
Q_{2\chi} \left[ g^{(m)} \right] ^2
\right\} / \left[ C^{(m+1)} \right] ^2 $ (crosses) 
on the 20-site lattice with $\epsilon = 0.02$.}
\label{fig:rat20b}
\end{center}
\end{figure}

\begin{figure}[h]
\begin{center}
\scalebox{0.5}{\includegraphics{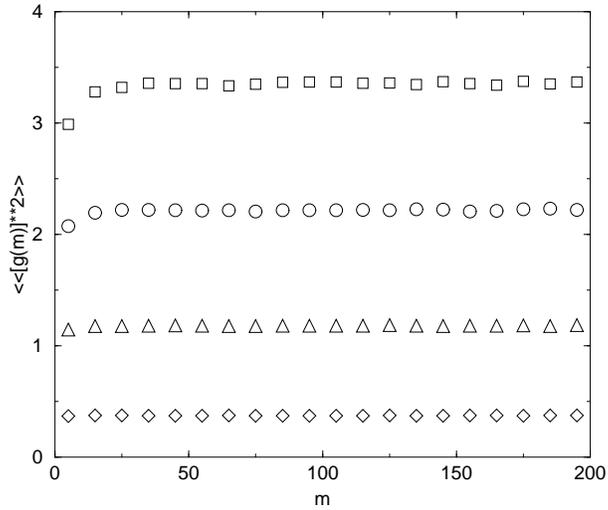}}
\caption{Statistical averages for $\left[g^{(m)}\right]^2$ on the 
20-site lattice as a function of $m$.
Squares, circles, triangles and diamonds show the data with 
$\epsilon = 0.02$, 0.015, 0.01 and 0.005, respectively.}
\label{fig:gg20}
\end{center}
\end{figure}

In this subsection we start from the exact eigenstate, namely, 
$\mid \psi^{(0)}\rangle =\mid \psi_{\rm E}\rangle $.
Then according to (\ref{psim}) 
we sequentially calculate $ \mid \psi^{(m)}\rangle $ up to $m=200$. 
Figs.~\ref{fig:wm20} - \ref{fig:gg20} present results on 
several quantities described in previous sections. 
All statistical averages in these Figures are calculated from 10 
samples for each value of $\epsilon$. 
We do not plot statistical errors because they are scarcely beyond 
the marks we used in the Figures.
    
In Fig.~\ref{fig:wm20} we present the results on $w^{(m)} 
\equiv \langle \psi_{\rm E} \mid \psi^{(m)} \rangle$ 
for several values of $\epsilon$ as a function of $m$.
We see that, for each value of $\epsilon$ less than or equal to 0.02, 
$w^{(m)}$ $(m > \ \sim 30)$ fluctuates around a finite value defined 
by $\epsilon$. This is the very evidence for the SSS equilibrium. 
While, when $\epsilon \geq 0.025$, $w^{(m)} $ rapidly decreases to 
zero as $m$ grows.
Therefore no SSS equilibrium takes place in this parameter region.
Fig.~\ref{fig:cm20} shows the normalization factor $C^{(m)}$ 
we calculate from (\ref{cmsquare}). 
Our results in Fig.~\ref{fig:cm20} confirm 
that values of $C^{(m)}$ are around the exact eigenvalue $Q$ 
when the SSS equilibrium occurs.

Figs.~\ref{fig:qq20}, \ref{fig:rat20a} and \ref{fig:rat20b} are to 
endorse relations (\ref{as1}), (\ref{as2}) and (\ref{wright}). 
In Fig.~\ref{fig:qq20} we plot values of 
$\langle \zeta ^{(m)} \mid \hat Q ^2 \mid \zeta ^{(m)} \rangle $
and $\langle \chi ^{(m)} \mid \hat Q ^2 \mid \chi ^{(m)} \rangle $ 
for some values of $\epsilon$.
It is clear that our assumptions (\ref{as1}), (\ref{as2}) are both acceptable 
because $\langle \zeta ^{(m)} \mid \hat Q ^2 \mid \zeta ^{(m)} \rangle $
and $\langle \chi ^{(m)} \mid \hat Q ^2 \mid \chi ^{(m)} \rangle $ 
hardly depend on $m$ when $m > \ \sim 20$.
We also observe that (\ref{as1}) and (\ref{as2}) hold for other values 
of $\epsilon$. 
The resultant value of $Q_{2\zeta}$ ($Q_{2\chi}$) first 
increases (decreases), then decreases (increases), 
as we lessen the value of $\epsilon$. 
The change for $Q_{2\chi}$ is, however, too small to read in this Figure. 
Figs.~\ref{fig:rat20a} and \ref{fig:rat20b} present ratios of 
three terms in the right-hand side of (\ref{wright}) to 
$\left[ C^{(m+1)} \right] ^2 $ for $\epsilon = 0.025$, 
where no equilibrium is observed, and for $\epsilon = 0.02$ with 
which the system realizes the equilibrium.
We see that the sum of the three ratios for each $m$ is 
in good agreement with the expected value, namely 1.0.  
We also observe such agreements for other values of $\epsilon$.
Therefore we can say that the relation (\ref{wright}) is acceptable 
regardless of the equilibrium.
In other words, we can conclude that the cross terms 
$2 Q^2 w^{(m)} \langle \psi_{\rm E} \mid \chi^{(m)} \rangle g^{(m)}$ and 
$2 s^{(m)} \langle \zeta^{(m)}  \mid \hat Q^2 \mid \chi^{(m)} 
\rangle g^{(m)} $ in (\ref{cmp1sq}) are always negligible for this system.  

Let us next examine the relation (\ref{weq0sol}) in the SSS equilibrium. 
Fig.~\ref{fig:gg20} plots 
$\langle \! \langle \left[g^{(m)}\right]^2 \rangle \! \rangle$ 
with $\epsilon =0.02$, 0.015, 0.01 and 0.005.
We see, in accordance with what we have assumed at the beginning 
of (\ref{g2uplim}), that fluctuations in $\left[g^{(m)}\right]^2$ 
due to different random number sequences are 
negligibly small and that $\left[g^{(m)}\right]^2$ 
for each value of $\epsilon$ becomes almost irrelevant to $m$ for 
$m > \ \sim 20$. 
For any value of $\epsilon$ shown in Fig.~\ref{fig:gg20} we observe that 
the value of $w^{({\rm eq})}$ estimated from 
$\sqrt{1-\langle \! \langle \ \langle \chi^{(m)} \mid 
\hat Q ^2 \mid \chi^{(m)} \rangle \  \rangle \! \rangle \ \langle \! 
\langle \left[g^{(m)} \right]^2 \rangle \! \rangle /
\left( Q^2 - \langle \! \langle \ \langle \zeta^{(m)} \mid \hat Q ^2 
\mid \zeta^{(m)} \rangle \ \rangle \! \rangle \right)}$ 
is in good agreement with values of $\langle \! \langle w^{(m)} 
\rangle \! \rangle$ in Fig.~\ref{fig:wm20}. 

Finally we comment on how many basis states should be included 
in our calculation.
As a result of the stochastic state selection, $N_{\rm a} ^{(m)}$, 
the number of non-zero coefficients in the expansion of 
$\hat M _{\{\eta^{(m+1)}\}} \mid \psi^{(m)} \rangle$, 
is much less than $N_{\rm b} ^{(m)}$, the number of non-zero coefficients 
in the expansion of $\mid \psi^{(m)} \rangle$. 
Hence it is not $N_{\rm a} ^{(m)}$ but $N_{\rm b} ^{(m)}$ 
that determines the necessary memory resources.
For each value of $\epsilon $ we observe that after several 
iterative operations to generate 
$\hat Q \hat M _{\{\eta^{(m)}\}} \mid \psi^{(m-1)} \rangle 
(= C^{(m)} \mid \psi^{(m)} \rangle )$,
$N_{\rm b} ^{(m)}$ does not increase anymore.
The data also show very small deviations for different random 
number sequences. 
Therefore the upper limit of $N_{\rm b} ^{(m)}$, which is dependent 
on the value of $\epsilon$, is critical for the numerical study. 
Some results for the 20-site system are as follows.  
When $\epsilon = 0.05$ the upper limit of $N_{\rm b} ^{(m)}$ is 
slightly less than $1 \times 10^5$, namely we can do with only 
one half of the $N_{\rm V} = 184,756$ basis states. 
With the value $\epsilon = 0.02$, for which the system exhibits the SSS 
equilibrium, we need about 80 $\%$ of the $N_{\rm V} $.  
The ratio $N_{\rm b} ^{(m)}/N_{\rm V}$ amounts to about 0.88 
if we employ the value $\epsilon = 0.01$.   
  
\subsection{$N_s$=32 and 36 Results}
\label{ns32and36}

Here we concentrate ourselves to the zero-momentum states which are even under 
rotations and reflections.  
Based on these conditions we construct a basis which possesses 
translation, rotation and reflection symmetries of the lattice \cite{szp}.
The total numbers of the $S_z = 0$ basis states are then $\sim 2.4 \times 10^6$
and $\sim 3.2 \times 10^7$ for the $N_{\rm s} = 32$ and 36 lattices, 
respectively. 
Since we can employ any initial trial state 
as far as it has some overlap with the exact eigenstate, 
we start from the N$\acute{\rm e}$el state for simplicity.
The value of $l$ in the operator $\hat Q = l\hat I - \hat H_{J_1 J_2}$ 
is set to be 4.8 (5.4) for the 32-site (36-site) lattice  
so that the largest eigenvalue for $\hat Q$ gives the lowest eigenvalue of 
$\hat H_{J_1 J_2}$. 

\begin{figure}[p]
\begin{center}
\scalebox{0.5}{\includegraphics{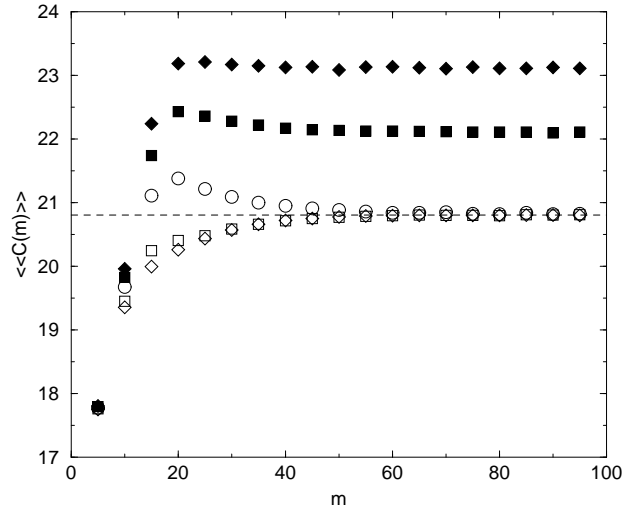}}
\caption{Statistical averages for $C^{(m)}$ on the 32-site lattice 
as a function of $m$, where each average is calculated from 30 samples.
Filled diamonds, filled squares, open circles, open squares and open 
diamonds show the data with $\epsilon = 0.015$, 0.013, 0.010, 0.005 
and 0.0025, respectively. 
The dashed line indicates the value $4.8 - E = 20.8031$ \cite{szp}. }
\label{fig:cm32}
\end{center}
\end{figure}

\begin{figure}[h]
\begin{center}
\scalebox{0.5}{\includegraphics{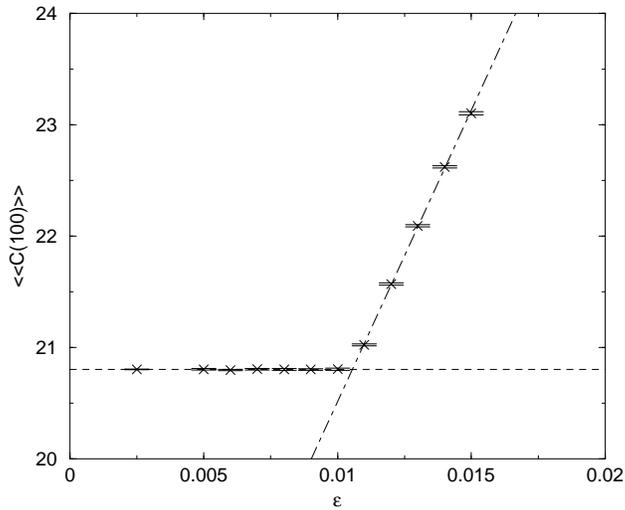}}
\caption{Statistical averages for $C^{(100)}$ on the 32-site lattice 
as a function of $\epsilon$. Each average is calculated from 30 samples. 
The dashed line indicates the value $4.8 - E = 20.8031$ \cite{szp} and 
the dot-dashed line shows the result 
from the linear fit for data with $ 0.01 < \epsilon \leq 0.015$.}
\label{fig:cme32}
\end{center}
\end{figure}

Let us first report results on the 32-site lattice.  
Changing values of $\epsilon$ we calculate 
$\langle \! \langle C^{(m)} \rangle \! \rangle$ 
($m \leq 100$) from 30 samples for each value of $\epsilon$. 
Figs.~\ref{fig:cm32} and ~\ref{fig:cme32} present the results, where 
statistical errors are so small that we omit them. 
In Fig.~\ref{fig:cm32} the results are shown as a function of $m$. 
We see that the data become almost constant when $m > \sim 50$. 
Fig.~\ref{fig:cme32} plots values of 
$\langle \! \langle C^{(100)} \rangle \! \rangle$ 
as a function of $\epsilon$.   
A linear decrease to meet a kink at $\epsilon \sim 0.01$ 
is observed in the Figure, 
and below this value the results are almost constant. 
We therefore come to a conclusion that the SSS equilibrium manifests 
itself when $\epsilon \leq 0.01$ and that 
measurements of the normalization factors $C^{(m)}$ in this parameter 
range enable us to estimate the lowest energy eigenvalue of the 
system \cite{footg}.
We observe that the estimated values from
$\langle \! \langle C^{(100)} \rangle \! \rangle$  
with $\epsilon = 0.01$ and $\epsilon = 0.0025$ are $E=-16.0052 \pm 0.0092$ 
and $E=-16.0047 \pm 0.0028$, respectively. These values are 
in good agreement with $E= -16.0031$, 
the exact ground-state energy obtained from ref. \cite{szp}.
The upper bound of $N_{\rm b}^{(m)}$ is $ \sim 1.7 \times 10^6$ for 
$\epsilon = 0.01$, $ \sim 1.8 \times 10^6$ for $\epsilon = 0.005$ 
and $\sim 1.9 \times 10^6$ for $\epsilon = 0.0025$, 
that is about $73 \%$, $77 \%$ and $81 \%$ of $N_{\rm V}$.

\begin{figure}[h]
\begin{center}
\scalebox{0.5}{\includegraphics{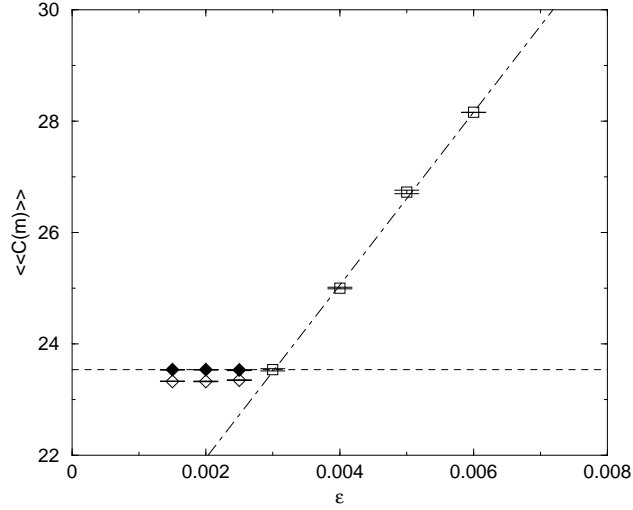}}
\caption{Statistical averages for $C^{(40)}$ (open squares, open diamonds) and 
$C^{(100)}$ (filled diamonds) on the 36-site lattice 
as a function of $\epsilon$. Each average for $\epsilon \leq 0.0025$ 
(the open diamond or the filled diamond)  
is calculated from 20 samples, while each averages for $\epsilon \geq 0.003$
(the open square) is calculated from 2 samples . 
The dashed line indicates the value of $5.4-E=23.5372$ \cite{szp}.
The dot-dashed line shows the result from the least square fit for data with 
$0.003 \leq \epsilon 
\leq 0.006$.}
\label{fig:cme36}
\end{center}
\end{figure}

Results on the 36-site lattice are qualitatively similar to those on 
the 32-site lattice.
In Fig.~\ref{fig:cme36} we show 
$\langle \! \langle C^{(40)} \rangle \! \rangle$ 
for $0.0015 \leq \epsilon \leq 0.006$ as well as 
$\langle \! \langle C^{(100)} \rangle \! \rangle$ 
for $0.0015 \leq \epsilon \leq 0.0025$. 
We observe a linear dependency of 
$\langle \! \langle C^{(40)} \rangle \! \rangle$ on $\epsilon$ 
in the range $0.003 \leq \epsilon \leq 0.006$, while data with 
less values of $\epsilon$ are almost constant. 
This kind of stability is observed with both
$\langle \! \langle C^{(40)} \rangle \! \rangle$ and 
$\langle \! \langle C^{(100)} \rangle \! \rangle$ for $\epsilon \leq 0.0025$, 
although the value of the constant is slightly different as we see 
in the Figure.     
From 20 samples of $\langle \! \langle C^{(100)} \rangle \! \rangle$ with 
$\epsilon = 0.0025$, 0.002 and 0.0015 we obtain $E=-18.1317 \pm 0.0089$, 
$-18.1357 \pm 0.0048$ 
and $-18.1364 \pm 0.0028$, respectively. Within the statistical errors 
they all agree with the known exact ground-state energy 
$E=-18.1372$ \cite{szp}.
The upper bound of $N_{\rm b}^{(m)}$ (ratio to the $N_{\rm V}$) is about 
$ 2.4 \times 10^7$ ($0.75$) when $\epsilon=0.0015$.   

\subsection{$N_s=40$ Results}
\label{ns40}

In this subsection we show results on a 40-site lattice. 
Since this lattice lacks the reflection symmetry, 
we employ a basis which possesses the up-down symmetry of
spins in addition to the translation and rotation
symmetries of the lattice.
The size of the full vector space with $S_z=0$ is then $N_{\rm V}
\simeq 4.3 \times 10^8$.
We define $\hat Q \equiv 6 \hat I - \hat{H} _{J_1 J_2}$ in this subsection.

\begin{figure}[h]
\begin{center}
\scalebox{0.5}{\includegraphics{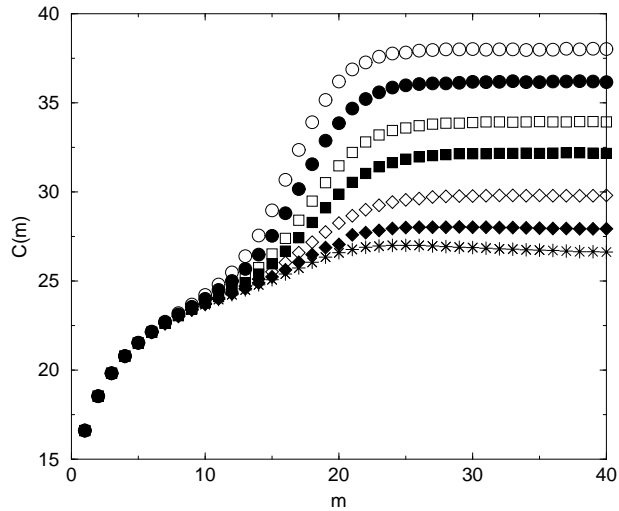}}
\caption{Values of $C^{(m)}$ on the 40-site lattice 
for $\epsilon \geq 0.001$ as a function of $m$. 
The initial trial state is the N$\acute{\rm e}$el state. 
Open circles, filled circles, open squares, filled squares, open
 diamonds, filled diamonds and asterisks denote data with 
$\epsilon$ = 0.005, 0.0035, 0.0025, 0.002, 0.0015, 0.0012 and 0.001, 
respectively.}
\label{fig:ifig_obo}
\end{center}
\end{figure}

For $\epsilon \geq 0.001$ we calculate $C^{(m)}$ starting from the 
N$\acute{\rm e}$el state. 
In the range $\epsilon \geq 0.002$ two samples are calculated 
for each value of $\epsilon$.
We observe that deviations of $C^{(m)}$ between different random number 
sequences are very small; the statistical errors are $0.5\%$ at most.
Since our data from one sample would have enough precision for this system,  
only one sample is calculated for each value of $\epsilon$ which is less
than 0.002.
Fig.~\ref{fig:ifig_obo} shows some of the results for $\epsilon \geq 0.001$,
where we present data from each sample as a function of $m$.
We see that $C^{(m)}$ ($m \geq 30$) for each value of $\epsilon$
becomes almost constant when $\epsilon \geq 0.0012$.
For $\epsilon = 0.001$, on the other hand, $C^{(m)}$ is still
decreasing at $m=40$. We therefore continue to calculate $C^{(m)}$ with 
$\epsilon =0.001$ up to $m=65$.
The values we obtain are $C^{(40)}=26.620$ and $C^{(65)}=26.417$. 

\begin{figure}[h]
\begin{center}
\scalebox{0.5}{\includegraphics{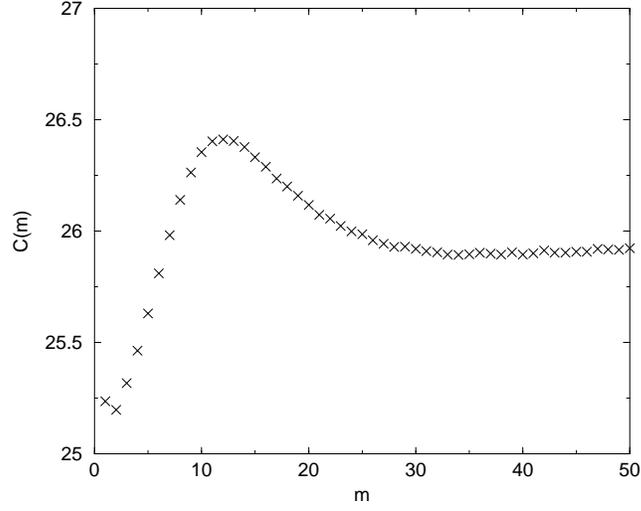}}
\caption{Values of $C^{(m)}$ on the 40-site lattice  
with $\epsilon = 0.0008$ as a function of $m$. 
The initial trial state is an approximate ground state which yields 
$\langle \psi ^{(0)} \mid \hat Q \mid \psi ^{(0)} \rangle =
\langle \psi ^{(0)} \mid 6 \hat I - \hat H \mid \psi ^{(0)} \rangle = 
18.814$. }
\label{fig:ifig_tmy}
\end{center}
\end{figure}

In further calculations of $C^{(m)}$ with $\epsilon = 0.0008$, we 
employ a better trial state which is constructed in the same manner as 
that in ref. \cite{mune5}. 
Values of $C^{(m)}$ thus obtained are presented in Fig.~\ref{fig:ifig_tmy}. 
The upper bound of $N_{\rm b}^{(m)}$ is $\sim 3.3 \times 10^8$ 
($76 \%$ of $N_{\rm V}$).
  
\begin{figure}[h]
\begin{center}
\scalebox{0.5}{\includegraphics{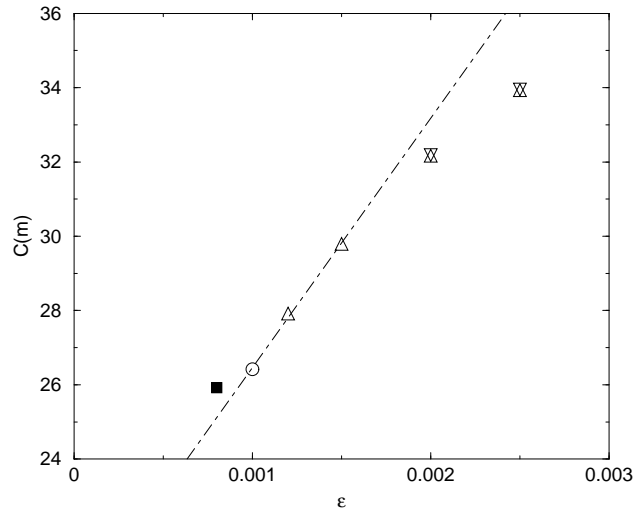}}
\caption{Values of $C^{(40)}$ (open triangles), $C^{(50)}$ 
(the filled square) and $C^{(65)}$ (the open circle) on the 40-site lattice 
as a function of $\epsilon$. Each datum is obtained from one sample.  
The dot-dashed line shows the result from the least square fit for data with 
$0.001 \leq \epsilon \leq 0.0015$. 
}
\label{fig:jfig_obo}
\end{center}
\end{figure}

In Fig.~\ref{fig:jfig_obo} we plot values of $C^{(40)}$ 
(for $0.001 < \epsilon \leq 0.0025$),
the value of $C^{(65)}$ (for $\epsilon = 0.001$) and 
the value of $C^{(50)}$ (for $\epsilon = 0.0008$) as a function 
of $\epsilon$.
We see that the datum with $\epsilon = 0.0008$ is located above 
the line in the Figure which is determined by data between 
$\epsilon = 0.001$ and $\epsilon = 0.0015$.   
Therefore, for the same reason as in the 32-site and 36-site cases,
we conclude that the 40-site system realizes the SSS equilibrium when  
$\epsilon = 0.0008$.

In order to evaluate the lowest energy of the system without any
statistical average, we attempt to find a zone which should include the 
true energy eigenvalue. 
The upper (lower) value of this zone is determined by the maximum 
(minimum) value of $C^{(m)}$ with $m \geq m_e$, 
where $m_e$ denotes a value of $m$ above which the system is in the SSS 
equilibrium.
This approach seems to work nicely in the 32-site and the 36-site
systems, resultant zones being much wider than the stripes obtained 
by the statistical treatments \cite{footzn}.  
Our best estimate for the zone in the 40-site system is 
$-19.92 \leq E \leq -19.89$, which we obtain from values of $C^{(m)}$ 
with $\epsilon = 0.0008$ and $m_e = 40 \leq m \leq 50$.

\section{Summary and Discussions}
\label{sumdis}

In this paper we study a frustrated quantum spin model using the SSS 
(Stochastic State Selection) method. Our purpose is to see whether the SSS 
equilibrium \cite{mune4}, a kind of equilibrium   
which we have found in a study of quantum spin models
with positive-definite Hamiltonians, is also realized in frustrated
systems.
 
In the SSS method we start from an initial trial state and 
recursively calculate the $m$-th normalized intermediate state 
$(m=1,2,3, \cdots)$. 
Each generating procedure is as follows.
First we operate the $m$-th random choice operator to the $(m-1)$-th 
normalized intermediate state. 
By this operation the effective size of the vector space is drastically 
reduced. The rate of the reduction is controlled by a parameter $\epsilon$. 
Then successively, we operate $\hat Q \equiv l \hat I - \hat H$ 
to the reduced state. This operation again increases 
the number of the basis states which are relevant to the resultant state. 
Finally we normalize the state to obtain the $m$-th
normalization factor and the $m$-th normalized intermediate state.
The system is in the SSS equilibrium if, 
after repeating the procedure many times, 
the $m$-th normalized intermediate state comes 
to contain a finite portion of the ground state and  
this portion is irrelevant to $m$.  
It should be kept in mind that in unfrustrated models 
the SSS equilibrium is observed for any value of the parameter $\epsilon$.

What we assert for frustrated quantum spin models is that the SSS
equilibrium is also realized in these models with {\em small} values of 
the parameter $\epsilon$. After analytical arguments in section \ref{exssseql} 
we present results from numerical calculations for the $J_1$-$J_2$
Heisenberg model on 20-, 32-, 36- and 40-site lattices with $J_2/J_1=0.5$.
We observe that these systems are in the SSS equilibrium in some range
of $\epsilon$. 
We also confirm that in the equilibrium we can effectively estimate 
the lowest energy eigenvalue of the system.
Our results on 20-, 32- and 36-site lattices are in good agreement with 
known exact ground-state energies. 
On a 40-site lattice
we obtain the result $-19.92 \leq E \leq -19.89$,
the basis states required for this calculation being maximally $76 \%$
of the full vector space. 
The energy per site is then $-0.4980 \leq E/N_{\rm s} \leq -0.4973$,  
which is greater than reported values for smaller lattices \cite{szp}, 
$E/N_{\rm s}=-0.503810$, $-0.500096$ and $-0.500615$ 
for $N_{\rm s}=36$, 32 and 20, respectively.
Since the value for $N_s=36$ seems to be irregular, 
we omit it from the extrapolation.
The extrapolation to $N_{\rm s} \rightarrow \infty$ using our 
$N_{\rm s}=20$, 32 and 40 data yields $-0.4993 \leq 
(E/N_{\rm s})_{\infty} \leq -0.4946$.

Several remarks are now in order.
First, let us emphasize the wide availability of the SSS method in numerical studies.  
As we have shown in previous works, 
the SSS method is applicable to various models with various symmetries 
and various system sizes, in frustrated cases as well as in unfrustrated ones.

Using this method it is possible to calculate many physical quantities which are 
given by expectation values of some operators.   
Essential points of the calculation are as follows. 
Suppose we want to calculate an expectation value 
$\langle \psi \mid \hat O \mid \psi \rangle$, where $\hat O$ is a given operator and 
$\mid \psi \rangle$ denotes either the exact eigenstate $\mid \psi_{\rm E} \rangle$ 
or an approximate state $\mid \psi_{\rm A} \rangle$.
In the SSS method we can generate a state $\mid \psi \rangle + \mid \chi \rangle g$
instead of $\mid \psi \rangle$, where $\mid \chi \rangle g$ is the stochastically 
determined random state. 
Since $\mid \psi \rangle + \mid \chi \rangle g$ is described by less number of 
non-zero coefficients compared to $\mid \psi \rangle$ itself, the method enables 
us to calculate $\left( \ \langle \psi \mid + g \langle \chi \mid  \ \right) \ \hat O 
 \ \left( \ \mid \psi \rangle + \mid \chi ' \rangle g' \ \right) $. 
Then all we should do in addition is to take a statistical average of this quantity, 
because a mathematical relation  
\begin{eqnarray*}
\langle \! \langle \ \left( \ \langle \psi \mid + g \langle \chi \mid  \ \right) \ \hat O 
 \ \left( \ \mid \psi \rangle + \mid \chi ' \rangle g' \ \right) \ \rangle \! \rangle 
= \langle \psi \mid \hat O \mid \psi \rangle  
\end{eqnarray*} 
follows from basic properties of the random state.
In ref. \cite{mune3}, for example, we calculate the static structure factor 
of the spin-1/2 anti-ferromagnetic Heisenberg model on a 36-site triangular 
lattice in addition to $S_z=0$ and $S_z=1$ energy eigenvalues.

The SSS method provides a fundamental improvement in Monte Carlo techniques. 
Therefore it is easily combined with lots of established techniques 
such as the power method \cite{mune,mune2,mune3} 
and the Lanczos method \cite{mune2,mune5}. 
A study of the stochastic diagonalization by H. De Raedt and M. Frick \cite{dera} 
suggests us that a combination of the SSS method with the Jacobi method would also 
be interesting.   

How large is the cluster size we can manage in calculations 
with the SSS method, then.
It of course depends on aims and models.
In study of the energy eigenvalue, for instance, our purpose in 
refs. \cite{mune, mune2, mune3, mune5} is to obtain upper bounds of eigenvalues
from approximate eigenstates. While in ref. \cite{mune4} and in this 
study we want to evaluate the eigenvalue itself through non-zero overlaps 
with the exact eigenstate which are realized in the SSS equilibrium. 
The system sizes of the latter case are generally smaller than those of the former 
one because smaller values of the parameter $\epsilon$ become necessary.

As for models, in addition to the $J_1$-$J_2$ model 
we have studied the spin-1/2 Heisenberg models 
on a square lattice up to 64 sites \cite{mune, mune4}, 
on a triangular lattice up to 48 sites \cite{mune5}, 
the Shastry-Sutherland model up to 64 sites \cite{mune2}.
We found that the method is especially effective for the Shastry-Sutherland model. 
This can be understood because of the very compactness of its low-lying eigenstates, 
where basis states with relatively large coefficients dominate if a restructured 
dimer-like basis \cite{restr} is employed. 

We conclude this article with a comment whether one can apply our method to 
other frustrated models. 
From theoretical points of view, the SSS equilibrium would be observed 
in most models because our discussion in section \ref{exssseql} is based 
on very moderate assumptions.
How small the parameter $\epsilon$ should be is, however, a model-dependent 
problem which remains to be clarified in future works.
If a system realizes the SSS equilibrium with moderate 
values of $\epsilon$ it is possible to study 
many physical properties of the system on large lattices.

\vskip 1cm\noindent
{\bf {\Large{\em Note added}}} \\

After submitting this paper for publication we became aware of the work by means of 
the exact diagonalization \cite{spinpack}, which suggests 
that our error estimate might need more examinations in future studies.

\vskip 1cm\noindent
{\bf {\Large Acknowledgement}} \\

We are grateful to Dr. J. Schulenburg for his information on the result from  
the exact diagonalization.


\begin{thebibliography}{99}
\bibitem{footsss}
The original SSS method used in refs. \cite{mune,mune2} is now replaced by the recursive 
SSS (RSSS) method which has been proposed in ref. \cite{mune3}.
Therefore we refer to the recursive SSS method as the SSS method in this paper.
\bibitem{footuf}
In a previous paper \cite{mune4} we state the model is non-frustrated when 
$J_2 / J_1 = 0$.  
In this paper we call both the $J_2/J_1 < 0 $ and the $J_2/J_1 = 0$ cases unfrustrated.
\bibitem{footeta}
Note that, as was discussed in ref. \cite{mune3}, 
this average should be calculated with a
fixed $\{ \{\eta^{(m)}\},\{\eta^{(m-1)}\}, \cdots, \{\eta^{(1)}\} \}$.
\bibitem{footequil}
This is the same equilibrium as we called the RSSS equilibrium in ref. \cite{mune4}.
\bibitem{footcond}
We observed that these conditions are fulfilled for any value of $\epsilon$ 
we have studied.
\bibitem{footsz}
Study with other symmetries is also possible. For example, we study both $S_z=0$ and 
$S_z=1$ sectors in refs. \cite{mune2,mune3}.
\bibitem{footg}
It should be kept in mind that all we can estimate here is, of course, 
the lowest energy eigenvalue in the sector which has the same symmetries as our basis.
\bibitem{footzn}
In the 36-site case, for example, the zone obtained from one sample 
with $\epsilon = 0.002$ and $m_e=60 \leq m \leq 100$ is 
$-18.1563 \leq E \leq -18.0964$. This includes a stripe 
$-18.1405 \leq E \leq -18.1309$  ($E= -18.1357 \pm 0.0048$), 
the value we have obtained in the previous subsection.
\end{thebibliography}
\end{document}